\documentclass[
  reprint,
  nofootinbib,
  amsmath,amssymb
]{revtex4-2}

\usepackage{hyperref}
\usepackage{graphicx}
\usepackage{dcolumn}
\usepackage{bm}

\usepackage{natbib}
\usepackage{amsmath}
\usepackage{amssymb}
\usepackage{hyperref}

\usepackage{algorithm}%
\usepackage{algorithmicx}%
\usepackage{amsfonts}
\usepackage{booktabs}
\usepackage{graphicx}
\usepackage{subcaption}
\usepackage{algpseudocode}%
\usepackage{url}
\usepackage{dsfont}
\usepackage{xcolor} 
\usepackage{xspace} 
\usepackage{dblfloatfix}

\newcommand{\pythia}{\texttt{Pythia8}\xspace}

\usepackage[shortlabels]{enumitem}

\newcommand{\professor}{\texttt{Professor}\xspace}
\newcommand{\apprentice}{\texttt{Apprentice}\xspace}
\newcommand{\rivet}{\texttt{Rivet}\xspace}
\newcommand{\yoda}{\texttt{YODA}\xspace}
\newcommand{\HEPMC}{\texttt{HEPMC}\xspace}
\newcommand{\GPytorch}{\texttt{GPytorch}\xspace}
\newcommand{\BOTorch}{\texttt{BOTORCH}\xspace}
\newcommand{\BOTORCH}{\texttt{BOTORCH}\xspace}
\newcommand{\PETRA}{\texttt{PETRA}\xspace}
\newcommand{\PEP}{\texttt{PEP}\xspace}
\newcommand{\ALEPH}{\texttt{ALEPH}\xspace}
\newcommand{\LEP}{\texttt{LEPI}\xspace}

\newcommand{\CMS}{\texttt{CMS}\xspace}
\newcommand{\CUET}{\texttt{CUET}\xspace}
\newcommand{\Monash}{\texttt{Monash}\xspace}
\newcommand{\LHAPDF}{\texttt{LHAPDF}\xspace}

\begin{document}

\preprint{AAPM/123-QED,FERMILAB-PUB-25-0323-CSAID}

\title{Bayesian Optimization of \pythia Tunes}

\author{Ali Al Kadhim}
  \email{aalkadhim@fsu.edu.}

\author{Harrison B Prosper}%
 \email{hprosper@fsu.edu}
\affiliation{Department of Physics, Florida State University, Tallahassee, USA}%

\author{Stephen Mrenna}
\email{mrenna@fnal.gov}
\affiliation{Computational Science and AI Directorate, Fermilab, Batavia, Illinois, USA}

\date{\today}


\begin{abstract}
    A new tune (set of model parameters) is found for the six most important parameters of the \pythia final state parton shower and hadronization model using Bayesian optimization. The tune fits the \LEP data from \ALEPH better than the default tune in \pythia. To the best of our knowledge, we present the most comprehensive application of Bayesian optimization to the tuning of a parton shower and hadronization model using the \LEP data.
\end{abstract}

\maketitle

\onecolumngrid
\twocolumngrid

\section{Introduction and Motivation}
\label{sec:Introduction_and_Motivation}
An understanding of confinement (for which there is a 1M dollar prize from the Clay Institute\footnote{\url{https://www.claymath.org/millennium/yang-mills-the-maths-gap/}}) requires
the identification of the mechanism by which colored partons become color singlet hadrons. This phenomenon is currently described by models that are inspired by the strong interaction, but are not
derived from first principles.

Motivated by evidence from QCD factorization theorems 
\cite{QCD_factorization_Collins}, 
it is assumed that hadronization is a universal process that can be factorized from the perturbative calculations. 
If this is true, the same hadronization model can be used both in proton-proton collisions as well as electron-positron collisions:
a model fit to \LEP electron-positron collision data can be applied to simulations at the \texttt{LHC}.
An advantage of studying hadronization using \LEP data is that absence of phenomena associated with the break-up of the colliding beams.

\LEP provided a nearly perfect environment to study hadronization effects 
\cite{ALEPH_1996_S3486095}, one characterized by a
well-defined initial $e^+ e^-$ state and a high event rate at the $Z$ mass 
\cite{DELPHI:1996sen}. 
Furthermore, a large improvement in the quality of the data and a reduction of systematic errors was achieved at \LEP with respect to what was achieved in the era of \PETRA and \PEP 
\cite{QCD_Event_Generators}. 
Finally, using one dataset from one experiment avoids the problem of having to correct for correlations that exist between different experiments and datasets.

Currently, two phenomenological models of hadronization\footnote{``Hadronization" and ``fragmentation" can be used almost interchangeably.} 
are widely studied: cluster hadronization 
\cite{ORIGINAL_CLUSTER_MODEL} 
and 
string fragmentation 
\cite{ANDERSSON198331}. 
The Lund string model 
\cite{ANDERSSON198331} 
(summarized in Sec.~\ref{sec:Lund_parameters}) is the basis for the hadronization model in \pythia.
This model will be the subject of the current study.
The task of studying the Lund model seems daunting, since it requires expensive simulations with many possible, different model parameters.
The development of efficient tools to support the tuning of these parameters is a high priority for such studies.

The choice of experimental data used to tune the Lund string model parameters is an important consideration. The experimental data should cover a wide range of physics so as not to miss
the full implications of the model. Surveys have been conducted which conclude that model parameters are best determined by fitting the model to inclusive distributions, jet rates and shape distributions simultaneously
\cite{DELPHI:1996sen}. 
For these reasons, we chose to use the \LEP data for QCD observables collected by the \ALEPH detector, published in Ref. 
\cite{ALEPH_1996_S3486095}, 
for the tuning of the Lund hadronization parameters. The data offer a rich set of observables to tune these parameters, and comprises a collection of $\mathcal{O}(30)$ event-shape variables such as thrust, sphericity, aplanarity, and inclusive distributions such as $p_T^{in}, p_T^{out}$ and the rapidity with respect to the thrust axis. See Ref.~\cite{ALEPH_1996_S3486095} 
for descriptions of these observables. 

The investigations reported in this paper have two goals:
\begin{enumerate}
  \item Investigate the use of Bayesian optimization for tuning \pythia parameters by minimizing an objective function. In principle, Bayesian optimization is ideally suited for such a problem because of the high computational cost of evaluating the objective function, which requires simulating a large number of events at each tune parameter point.
  \item Assess whether Bayesian‐optimized tuning converges to the same parameter point as obtained with methods such as \professor~\cite{professor} or \apprentice~\cite{krishnamoorthy_apprentice_2021}.
\end{enumerate}

\bigskip

The remainder of this paper is organized as follows. 
In Sec.~\ref{Background_other_MC_tune_methods} we briefly review tuning methods and how our approach differs from them. In Sec.~\ref{sec:Lund_parameters} the six \pythia parameters, which are the focus of the tuning, are reviewed. Section~ \ref{sec:lep_data_event_simulation} briefly describes the \LEP event datasets employed and the objective function used in the tuning. Section~\ref{sec:Tuning_with_Bayesian_Optimization} describes implementation details of the Bayesian optimization framework.
The results of the Bayesian optimization of the six tune parameters are presented in Sec.~\ref{sec:Results} and  compared with the tune results from \professor/\apprentice‐based fits. Finally, Sec.~\ref{sec:conclusions} provides our conclusions. The interested reader can follow the directions in Sec.~\ref{sec:reproducibility} to reproduce all results presented in this paper.

\section{Monte Carlo tune methods}
\label{Background_other_MC_tune_methods}

\subsection{\professor and \professor-like tunes}

\professor \cite{professor} is the main event generator tuning method used by \CMS, \texttt{ATLAS} and other large collaborations. Recently, new tools for event generator tuning have become available, including \apprentice 
\cite{krishnamoorthy_apprentice_2021}, \texttt{Autotunes} 
\cite{autotunes}, 
and machine learning (ML) methods such as \texttt{MLHAD} 
\cite{MLHAD, Modelling_had_with_ML} 
and \texttt{HAD-ML} 
\cite{NNs_for_Reweighting_Tuning}. 
 While \professor (and \professor-like tuning methods such as \apprentice) have produced excellent results there is still room for improvement:
\begin{itemize}
    \item the estimated uncertainties are based on the Hessian inverse, consequently, there is no guarantee that the confidence sets have valid coverage;
    \item the polynomial fit to the Monte Carlo (MC)-based predictions is adequate for well-behaved distributions, but may fail for more singular distributions, (though see
    \cite{Austin:2019uci} for a rational polynomial fitter available in \apprentice) and
    \item the choice of the weights in these methods is done manually (though see \cite{Wang:2021gdl} for algorithms to select these weights).
\end{itemize}

\subsection{\CMS \CUET Tunes and MPI considerations}
\CMS has determined tunes, dubbed \CUET \cite{CMS_CUET}, which focus on the modeling of multi-parton interactions (MPI) in event generators.  Multi-parton interactions  are  $2 \rightarrow 2$ parton scatterings that occur within the same hadron-hadron collision, but are softer than the hard scattering ($p_T <3$ GeV). Like the hadronization models, MPI models are also  phenomenological with free parameters determined by fits to data.
\CMS has recently evolved beyond this and now uses the CPX family of tunes that use NLL $\alpha_s$ and NNLO parton distribution functions (PDF)~\cite{CMS:2019csb}.

\subsubsection{Tuning with Bayesian Optimization}
\label{Methods_Bayesian_Optimization}

 The tuning of an event generator involves generating a sufficient number of events at any given tune point, which is computationally costly. Optimization of a costly objective function is the kind of problem for which Bayesian optimization \cite{BayesOptOverview} is, in principle, well suited.
Bayesian optimization (BayesOpt) has been attempted before \cite{Ilten:2016csi} to tune event generators, but this paper presents a more thorough investigation of BayesOpt and its application to the tuning of \pythia.  Since BayesOpt remains unfamiliar to many particle physicists, Sec.~\ref{sec:Tuning_with_Bayesian_Optimization} presents the necessary background before the tune results obtained for \pythia are presented in Sec.~\ref{sec:Results}.

\section{Lund Fragmentation Model Parameters}
\label{sec:Lund_parameters}
The central assumption of the Lund model is that a quark-antiquark pair is subject to the QCD potential $V_{\mathrm{QCD}}(r) \approx -\frac{4}{3} \frac{\alpha_{\mathrm{s}}}{r}+\kappa r$, where $r$ is the distance between the quark and the antiquark and $\alpha_s$ is the strong coupling parameter. For an energetic quark and antiquark pair traveling in opposite directions, the linear term is most relevant.
The Lund model views the confining force between colored charges as a flux tube or a string with string tension $\kappa$. 
    The parameter $\kappa \approx 1$ GeV/fm was first inferred from Regge trajectories
\cite{REGGE_TRAJ}, 
and was later confirmed by lattice QCD calculations 
\cite{Lattice_QCD_confirm_kappa}.  
In the model, the string breaks through the quantum mechanical tunneling of quark-antiquark or diquark-antidiquark pairs out of the
vacuum.

\begin{figure}[htb]
    \centering
    \includegraphics[width=\columnwidth]{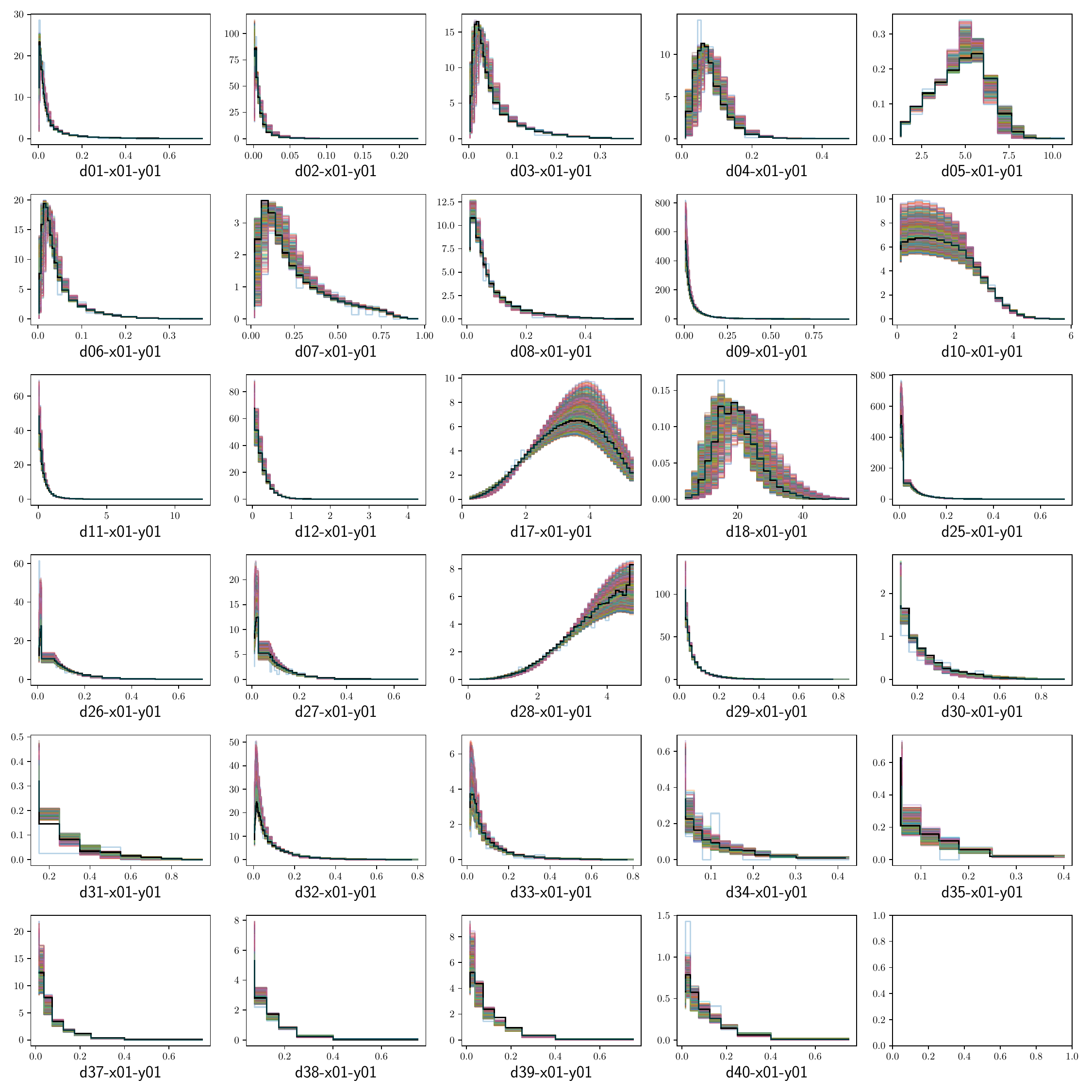}
    \caption{Impact of randomly sampling aLund and bLund (keeping the other parameters fixed to \Monash) and the associated effect it has on the \ALEPH distributions. Black lines show the experimental ALEPH data; colored curves correspond to model predictions for each random parameter draw.}
    \label{fig:varying_a_b}
\end{figure}

The Lund  fragmentation function, 
\begin{equation}
    f_{\text{Lund}}(z) \propto \frac{(1-z)^a}{z} \exp \left(-\frac{b m_{T}^2}{z}\right),
\label{lund_frag_general}
\end{equation}
is the probability density that a hadron $h$ with transverse mass 
\begin{equation}
    m_{T}^2=m^2+p_{T}^2
\label{m_perp}
\end{equation} 
is produced with lightcone momentum fraction $z$ 
\cite{Pythia_8.3},
 with the remainder, $1-z$, taken by the string. 
A large value of the $a$ parameter suppresses the hard region $z \rightarrow 1$ while a low value suppresses soft regions $z \rightarrow 0$. The $a$ and $b$ parameters are implemented in \pythia as \verb|StringZ:aLund| and \verb|StringZ:bLund|, respectively. The choice of \verb|StringZ:aLund| and \verb|StringZ:bLund| strongly impacts QCD observables (see, for example, Fig. \ref{fig:varying_a_b}).

While the $b$ parameter is universal, the
$a$ parameter does not need to be same for all flavors of quarks or diquarks.
The more general parameterization of fragmentation from parton $i$ to parton $j$ is:
\begin{align}
    f_{\text{Lund}}(z) & \propto \frac{1}{z^{1+r_Q b m_Q^2}}z^{\delta a_i-\delta a_j}( 1-z)^{\delta a_j} \nonumber\\ &\times \exp \left(-\frac{b m_{T}^2}{z}\right),
\label{lund_frag_flavor}
\end{align}
where we have included the  Bowler modification 
for massive endpoint quarks with mass $m_Q$: $r_Q=0$ unless $Q=c$ or $Q=b$. This introduces two new parameters \verb|StringZ:rFactC| and \verb|StringZ:rFactB| for the b and c quarks, respectively. 
The parameters $\delta a_i$  are only implemented for
strange quarks and diquarks, 
 and are denoted by \verb|StringZ:aExtraDiquark| and \verb|StringZ:aExtraSQuark|, respectively.

The tunneling of the partons from the vacuum imparts $p_T$ to the produced hadrons:  this is modeled using a Gaussian distribution in $p_x$ and $p_y$,
\begin{equation}
    P\left(p_x, p_y, \sigma_{T}\right)=\frac{1}{2 \pi \sigma_{T}^2} \exp \left[-\left(p_x^2+p_y^2\right) /\left(2 \sigma_{T}^2\right)\right],
\end{equation}
where the standard deviation, $\sigma_{T}$, is a free parameter denoted by
 \verb|StringPT:sigma| in \pythia.

The fragmentation model also accounts for strangeness suppression, that is, the creation of $s \bar{s}$ relative to $u \bar{u}$ and $d \bar{d}$ and is configured with the \pythia parameter \verb|StringFlav:probStoUD|. 
Also included in the tune is \verb|StringFlav:probQQtoQ|, the parameter that controls the degree of suppression of diquark production relative to quark production.

In \pythia the amount of QCD radiation in the parton shower depends strongly on the choice of $\alpha_S(M_Z^2)$, which is denoted by \verb|TimeShower:alphaSvalue|. The default value for this parameter was determined by a relatively crude tuning to \LEP data, which is why we choose to tune it. Finally, since the radiation rate  diverges as $p_T\rightarrow 0$, a regime in which perturbation theory is invalid, a cutoff is introduced, \verb|TimeShower:pTmin|, below which no QCD emissions are allowed. This value is usually chosen to be close to $\Lambda_{\text{QCD}}$ in order to have a smooth transition between low-$p_T$ perturbative emissions and non-perturbative string fragmentation.
The parameters tuned in this study are listed in Table \ref{table:pythia_param_table}.
Obviously, more parameters could be included, but we have identified a few critical ones
for the purpose of studying the Bayesian optimization methodology.


\section{\LEP Data and Event Simulation}
\label{sec:lep_data_event_simulation}
In this study \LEP electron-positron collision data, which are provided as histograms, are used to tune the parameters in Table \ref{table:pythia_param_table}. The \ALEPH histograms used and their labels are listed in Table \ref{table:all_hists}. The data in the $i^\text{th}$ bin is a measured cross section, $\sigma_i$, and its associated uncertainty, $\delta \sigma_{\text{data},i}$.
Ideally, the statistical uncertainty $\delta \sigma_{\text{MC}, i}$  in the associated cross section prediction $\sigma_{\text{MC}, i}$ should  $\ll \delta \sigma_{\text{data},i}$  so that $\delta \sigma_{\text{MC}, i}$ can be neglected.
In the \ALEPH histograms
the effective count, $n \approx \sum_{i} (\sigma_i / \delta \sigma_i)^2$, varies from $\sim 330$ to $\sim 950,000$. Therefore, in principle, at any given point in the tune space we should simulate $\gg 10^6$ events. However, for our current studies, 250,000 events per tune point are simulated, which yields predictions with a statistical precision that roughly match the overall precision of the data. The statistical uncertainties in the predictions are accounted for by adding $\delta \sigma_{\text{data},i}$ and $\delta \sigma_{\text{MC}, i}$ in quadrature to yield a cross section  standard deviation $\delta \sigma_i$.
For bins with standard deviations less than $10^{-3} \text{pb}$ the standard deviation is set to 1 to mitigate the danger that one bin dominates the objective function. 
Finally, histograms where the total cross section is $ < 0.1\,\text{pb}$  are omitted.

Since the correlations between histograms are not available, the objective function, $O(\mathbf{x})$, is taken to be a linear sum of $\chi^2$-like variates,
\begin{equation}
        O(\mathbf{x}) =  \sqrt{\frac{1}{N} \sum_\text{histograms}  \sum_{i=1}^{N_\text{bins}} \left( \frac{D_i - T_i(\mathbf{x})}{\delta\sigma_i} \right)^2}, 
        \label{eq:Ox}
\end{equation}
where $N= \sum_{\text{histograms}} N_\text{bins}$ is the total number of bins across all histograms,
 $D_i$ is the measured cross section in bin $i$, and $T_i$ is the theoretical cross section prediction  obtained via Monte Carlo simulation with \pythia for a given tune point $\mathbf{x}$. The normalization by the total number of bins and the square root in Eq.\,(\ref{eq:Ox}) reduces the dynamic range of $O(\mathbf{x})$. However, in our studies reducing the dynamic range of the objective function does not play a significant role.


.

\section{Tuning with Bayesian Optimization}
\label{sec:Tuning_with_Bayesian_Optimization}
 
Given an objective function $O: \mathcal{X} \rightarrow \mathbb{R}$ defined on a compact domain $\mathcal{X} \subseteq \mathbb{R}^d$, the goal of Bayesian optimization is to search the domain as efficiently as possible for a good approximation to the location of the global minimum
\begin{equation}
\label{Eq:minimization}
    \mathbf{x}_0 = \arg\min_{\mathbf{x} \in \mathcal{X}} O(\mathbf{x}) .
\end{equation}
Evaluating 
$O(\mathbf{x})$, however, is expensive because it requires simulating a large number of events on-the-fly.  Moreover, one does not have a closed form for the dependence of $O(\mathbf{x})$ on $\mathbf{x}$. But since the objective function can be computed at any given point it is possible to obtain ``observations" of the function at a set of sampled points. Bayesian optimization reduces a difficult, expensive  optimization problem into a sequence of less costly optimization problems in which a \emph{model} of $O(\mathbf{x})$ is optimized.  BayesOpt
(also known as \emph{surrogate-based optimization} 
\cite{booker_rigorous_nodate}) 
is especially useful for optimizing objective functions $O(\mathbf{x})$ for which  
there is no efficient method for estimating their gradients and/or if $O(\mathbf{x})$ is non-convex.
The only assumption about $O(\mathbf{x})$ in BayesOpt is that the objective is Lipschitz-continuous, that is, there exists some constant $C$, such that for all $\mathbf{x}_1, \mathbf{x}_2 \in \mathcal{X}$:
\begin{equation}
    \left\|O\left(\mathbf{x}_1\right)-O\left(\mathbf{x}_2\right)\right\| \leq C\left\|\mathbf{x}_1-\mathbf{x}_2\right\|.
\end{equation}

 
In principle, any sufficiently flexible function class can be used for the surrogate model, but in practice one uses  Gaussian processes (GPs) \cite{GP_book} because of their mathematical convenience.  Given the data $\mathcal{D} = \{X, Y\} = \{(\mathbf{x}_i, y_i) \}_{i=1}^n$, where $y_i = O(\mathbf{x}_i)$ is the observed value of the objective function at the tune parameter point $\mathbf{x}_i$ one first computes the posterior density, $p\left(f \mid \mathcal{D} \right)$, over surrogate models, $f(\mathbf{x})$, of the  objective function $O(\mathbf{x})$,
\begin{align}
    p\left(f \mid \mathcal{D}\right) & = \frac{p\left(\mathcal{D} \mid f\right) \, \pi(f)}{p(\mathcal{D})} , \nonumber\\
& = \frac{p\left(Y \mid \mathbf{X}, f\right)\, p(\mathbf{X} \mid  f) \, \pi(f)}{p(Y \mid \mathbf{X}  )\,p(\mathbf{X})} , \nonumber\\
& = \frac{p\left(y \mid \mathbf{X}, f\right) \, \pi(f)}{p(Y \mid \mathbf{X}  )} ,
    \label{eq:pfD}
\end{align}
where $\pi(f)$ is a prior over \emph{functions} modeled with a GP and where we have assumed that $p(\mathbf{X} \mid  f) = p(\mathbf{X})$. The quantity $p\left( Y \mid \mathbf{X}, f\right)$ is the likelihood function for the data $\mathcal{D}$. Next the posterior density together with another function, called the \emph{acquisition} function, is used to decide where the next observation of the objective function, $O(\mathbf{x})$, should be made such that one progressively arrives at a good estimate of the location, $\mathbf{x}_0$, of the global minimum. 

In the following, we consider how each of the three terms in Eq.\,(\ref{eq:pfD}) is computed and follow with a description of the BayesOpt algorithm noting which hyperparameters are most  relevant for deriving tunes.

\subsection{Gaussian Process Prior, $\pi(f)$}

A Gaussian process in $f(\mathbf{x})$, 
\begin{equation}
\label{eq:p_f}
    \mathcal{G} \mathcal{P}(f ; \mu, \mathbf{K}) ,
\end{equation}
is a set of random variables, any countable  set of which, $f(\mathbf{x}_1), f(\mathbf{x}_2),\cdots$. follow a multivariate normal distribution, where $\mu$ is the
\emph{mean function}
 given by
\begin{equation}
    \mu(\mathbf{x})= \mathbb{E}_{f\sim \mathcal{GP}}[f(\mathbf{x}) \mid \mathbf{x}] ,
\end{equation}
and $\mathbf{K}$  is the \emph{covariance function} (or kernel, or kernel function) whose matrix elements are given by
\begin{equation}
\begin{split}
k(\mathbf{x},\mathbf{x}')
  &= \mathrm{cov}\bigl[f(\mathbf{x}),f(\mathbf{x}') \mid \mathbf{x},\mathbf{x}'\bigr]\\
  &= \mathbb{E}_{f\sim\mathcal{GP}}\Bigl[\bigl(f(\mathbf{x})-\mu(\mathbf{x})\bigr)
      \bigl(f(\mathbf{x}')-\mu(\mathbf{x}')\bigr)\Bigr].
\end{split}
\label{eq:cov_def}
\end{equation}
The mean function for the prior is chosen to be $\mu(\mathbf{x}) = 0$, that is,
\begin{equation}
    \pi(f) = \mathcal{GP}(f; 0, \mathbf{K}).
    \label{eq:prior}
\end{equation}
For the matrix elements of the covariance function we choose the ARD Matern kernel~\cite{Matern_kernel}

\begin{align}
k(\mathbf{x}_i,\mathbf{x}_j)
  & =  C \ \frac{1}{2^{\nu-1}\,\Gamma(\nu)}
     \bigl(2\sqrt{\nu} d \bigr)^{\!\nu} \,
   H_{\nu}\!\bigl(2\sqrt{\nu}\ d \bigr) , \nonumber\\
 \text{with }   d &= (\mathbf{x_i} - \mathbf{x_j})^T S^{-1} (\mathbf{x_i} - \mathbf{x_j}),
\label{eq:matern_general}
\end{align}
where $S^{-1}$ is a diagonal matrix composed of the lengthscale parameter for each dimension, $C$ is an overall scale factor, and $\Gamma (\cdot)$ and $H_\nu (\cdot)$ are the gamma function and Bessel function of order $\nu$, respectively. The parameter $\nu$---which determines the smoothness of the functions $f(\mathbf{x})$---is set to $\nu=5/2$, which has been singled out as a good choice.

\subsection{Likelihood function, $p(Y \mid \mathbf{X}, f)$}



Suppose we have observed (that is, evaluated) the objective function $y = O(\mathbf{x})$ at  parameter points $\mathbf{X} = \mathbf{x}_1,\cdots,\mathbf{x}_n$. For the Bayesian inference to proceed, we need an expression for the likelihood function of the observations, $Y = y_1,\cdots, y_n$, where $y_i = O(\mathbf{x}_i)$. A typical assumption is  that 
\begin{equation}
    y_i = f(\mathbf{x}_i) + \varepsilon,
\label{Eq:noise_model}
\end{equation}
where $\varepsilon \sim \mathcal{N}(0, \sigma_\text{noise}^2)$ is Gaussian noise, yielding a diagonal multivariate Gaussian likelihood 
\begin{equation}
p(Y \mid \mathbf{X},  f) = \prod_{i=1}^n p(y_i \mid \mathbf{x}_i, f(\mathbf{x}_i) ). 
\label{eq:likelihood}
\end{equation}
This choice for the likelihood is particularly convenient because it yields a posterior density that is a Gaussian process with its mean and covariance functions available in closed form, as noted in the next section.

\subsection{Posterior density, $p(f \mid \mathcal{D})$}
Consider the set of observations $Y$ at the points $\mathbf{X}$ and another set of points $\mathbf{X}^* = \mathbf{x}^*_1,\cdots,\mathbf{x}^*_m$ at which we wish to estimate the objective function. A Gaussian process prior and a Gaussian likelihood yield a posterior density that is a Gaussian process 
with mean and covariance functions given by 
\begin{align}
    \mu(\mathbf{x}^*|\mathcal{D}) &=  \mu(\mathbf{x}^*) + 
    \mathbf{K}^T_{\mathbf{X} \mathbf{x}^*} \mathbf{K}_{\mathbf{X X}}^{-1} [Y - \mu(\mathbf{X})] 
    \label{eq:meanupdate}
\end{align}
and
\begin{align}
    k(\mathbf{x}*, \mathbf{x}^{*\prime}|\mathcal{D}) & = k(\mathbf{x}^*, \mathbf{x}^{*\prime}) - \mathbf{K}^T_{\mathbf{X} \mathbf{x}^*} \mathbf{K}_{\mathbf{X X}}^{-1}\mathbf{K}_{\mathbf{X} \mathbf{x}^*}, 
    \label{eq:covupdate}
\end{align}
respectively, where the prior mean, $\mu(\mathbf{x}) = 0$ (see Eq.\,(\ref{eq:prior}). The elements of the matrices $\mathbf{K}_{\mathbf{X} \mathbf{x}^*}$ and $\mathbf{K}_{\mathbf{X X}}$ are given by $k(\mathbf{x}_i, \mathbf{x}^*)$ and $k(\mathbf{x}_i, \mathbf{x}^\prime_j) + \sigma^2_\text{noise} \delta_{ij}$, $i, j = 1,\cdots,n$, respectively, where the kernel is given by Eq.\,(\ref{eq:matern_general}) and the likelihood by Eq.\,(\ref{eq:likelihood}). As more observations are added to the existing set the mean and covariance functions are updated recursively according to Eqs.\,(\ref{eq:meanupdate}) and (\ref{eq:covupdate}).

As in all Bayesian calculations, the posterior density completes the inference. However, one typically computes a few useful functionals of the posterior density including the predictive distribution
\begin{equation}
    p(y | \mathbf{x}^*, \mathcal{D}) = \int p(y \mid \mathbf{x}^*, f)  \, p(f \mid \mathcal{D}) \, df ,
\end{equation}
which in this context is the probability density that an observation of the objective function at the new point $\mathbf{x}^*$ yields the value $y$. The mean of the predictive distribution is often taken to be an estimate of the objective function at $\mathbf{x}^*$, while the width of the distribution is a measure of the uncertainty associated with that estimate.

\subsection{Acquisition functions}
\label{sec:acq_func}
Another important quantity that can be computed from the posterior density is the expected utility, $\mathbb{E}[u]$, of moving to the point $\mathbf{x}^*$ with the goal of moving along an efficiently selected sequence of points to the minimum of the objective function. The expected utility in this context is called an acquisition function of which several were considered in this study. The \emph{expected improvement} acquisition function ($\alpha_\text{EI}$) is defined by the utility function
\begin{equation}
\label{Eq:EI_util}
    u(f^*, y^+) = \max(y^+ - \xi - f^*, 0),
\end{equation}
where $f^*$ is the value of the surrogate function at 
point $\mathbf{x}^*$
and $y^+ = O(\mathbf{x}^+)$ is the lowest value of the objective function value found so far. The hyperparameter $\xi$ controls the degree to which the expected improvement favors exploration versus exploitation. Large values of $\xi$ penalizes exploitation in favor of exploration. 
For a GP the expected improvement can also be obtained in closed form
\cite{Jones1998EfficientGO} and is given by
\begin{equation}
\label{eq:EI}
\alpha_{\text{EI}}(\mathbf{x})
= \begin{cases}
  \begin{aligned}
    &\bigl(y^+ - \mu(\mathbf{x})-\xi\bigr)\,
      \Phi\!\bigl(\tfrac{y^+ -\mu(\mathbf{x})-\xi}{\sigma(\mathbf{x})}\bigr)\\
    &\quad +\,\sigma(\mathbf{x})\,
      \phi\!\bigl(\tfrac{y^+-\mu(\mathbf{x})-\xi}{\sigma(\mathbf{x})}\bigr),
  \end{aligned}
  & \sigma(\mathbf{x})>0,\\[8pt]
  0, & \sigma(\mathbf{x})=0.
\end{cases}
\end{equation}
where $\Phi$ is the cumulative distribution function (cdf) of the standard normal whose density is denoted by $\phi(\cdot)$, and $\mu(\mathbf{x})$ and $\sigma(\mathbf{x})$ are respectively the posterior mean and standard deviation of the GP at $\mathbf{x}$.
In Ref.\,\cite{10.5555/1626686} it is found that $\xi=0.01$ gives the best results in all settings, while a schedule for $\xi$ that encourages exploration early and exploitation later does \emph{not} work in practice, contrary to intuition. We therefore keep $\xi=0.01$ constant throughout our studies.

 The next point at which to make an observation of the objective function is found by maximizing the acquisition function:
\begin{equation}
\label{eq:x_next}
    \mathbf{x}^*=\mathop{\mathrm{arg\,max}}_{\mathbf{x}^{*} \in \mathcal{X}} \alpha\left(\mathbf{x}^*\right).
\end{equation}

\subsection{Optimizing the kernel and likelihood hyperparameters}
\label{sec:kernel_hyperparams}
The likelihood, Eq.\,(\ref{eq:likelihood}), and the prior, Eq.\,(\ref{eq:prior}), both contain hyperparameters, denoted collectively by $\phi = C, S^{-1}, \sigma_\text{noise}$, which need to be optimized. In a purely Bayesian approach the hyperparameters would be eliminated from the problem through marginalization with respect to a hyperprior. However, it is simpler to use an empirical Bayes \cite{empirical_bayes} approach and optimize the hyperparameters by maximizing 
\begin{equation}
    p(Y | \mathbf{X}, \phi) = \int p(Y \mid \mathbf{X}, f)  \, \pi(f) \, df ,
\end{equation}
that is, the prior predictive distribution with respect
to $\phi$
\begin{equation}
\label{eq:phi_MLE} 
    \hat{\phi} = \arg\max_{\phi} p(Y | \mathbf{X}, \phi) .
\end{equation}
The data, $\mathcal{D}_\text{Sobol} = \{Y, \mathbf{X}\}$, for the likelihood are observations, $Y$, of the objective function at a Sobol \cite{SOBOL196786} sequence of points $\mathbf{X}$ that provide a quasi-random sampling of the tune space.
Eq.\,(\ref{eq:phi_MLE}) is implemented using the \texttt{Adam}~\cite{kingma2014adam} optimizer.

\subsection{BayesOpt Algorithm}
Algorithm \ref{BayesOpt_algorithm} shows the key steps of the BayesOpt algorithm,  has its own set of hyperparameters or choices that need to be optimized to obtain satisfactory algorithmic behavior. 
\begin{algorithm}[H]
\caption{Bayesian Optimization with Expected Improvement}
\begin{algorithmic}[1]
\Statex \textbf{Require:}
\begin{enumerate}[a)]
    \item Number of BayesOpt iterations: $N_\text{BayesOpt}$.  
    \item Number of Sobol points: $N_\text{Sobol}$.
    \item Number of restarts: $N_\text{restarts}$.
\end{enumerate}
\Statex \textbf{Ensure:}
\Statex Estimate $\mathbf{\hat{x}}$ of global minimum $\mathbf{x}_0$
\State {\bfseries Initialize: $\mathcal{D} \gets \emptyset $}
\State Generate Sobol points $\{\mathbf{x}_i\}_{i=1}^{N_\text{Sobol}} \subset\mathcal{X}$
\State Observe $y_i = O(\mathbf{x}_i)$
\State $\mathcal{D} \gets\{(\mathbf{x}_i,y_i)\}_{i=1}^n$
\State {\bfseries Optimize hyperparameters $\phi$:}
    $$\phi^*=\arg\max_\phi\; p(Y|\mathbf{X},\phi)$$
\For{$t=1,\dots, N_\text{BayesOpt}$}
  \State {\bfseries Compute posterior functions:} 
  $\mu(\mathbf{x^*}),k (x^*, x^{*'})$
  \State {\bfseries Compute expected improvement $\alpha_\text{EI}(x^*)$} 
  \State {\bfseries Select next point:}
    $$\mathbf{x^*} = \arg\max_{\mathbf{x^*}\in\mathcal{X}} \alpha_\text{EI}(\mathbf{x^*})$$
  \State {\bfseries Observe:}
    $y = O(\mathbf{x}^*)$
  \State {\bfseries Augment data:}
    $\mathcal{D}\gets \mathcal{D}\cup\{(\mathbf{x},y)\}$
\EndFor
\State \Return $\displaystyle \mathbf{\hat{x}} = \arg\min_{\mathbf{x}} \mathcal{D}$
\end{algorithmic}
\label{BayesOpt_algorithm}
\end{algorithm}
\noindent
These hyperparameters are in addition to those of the kernel and likelihood function.
The following were found to be the key BayesOpt hyperparameters to be optimized and the key choices to be made:
\begin{enumerate}
    \item number of BayesOpt iterations, $N_\text{BayesOpt}$;
    \item number of Sobol points $N_\text{Sobol}$;
    \item optimization algorithm for the acquisition function, and
    \item the number of restarts in the optimization of the acquisition function, $N_\text{restarts}$.
\end{enumerate}

In order to cross-check our results, two different toolkits—\GPytorch \cite{gardner2018gpytorch} and \texttt{BOTORCH} \cite{balandat_botorch_2020}—are employed to carry out the Bayesian‐optimization studies, in three stages. In \textbf{stage 1},  an \emph{ab initio} implementation of BayesOpt was developed using \GPytorch with its hyperparameters optimized on toy objective functions defined on the \pythia tune parameter space. In \textbf{stage 2}, the BayesOpt was applied to \pythia using the choices and hyperparameter settings established in stage 1 to obtain a \pythia tune. In \textbf{stage 3}, a \pythia tune was obtained using the  built-in BayesOpt routines of \texttt{BOTORCH}. The \texttt{BOTORCH} implementation used a different acquisition function from that used in our \GPytorch implementation, namely,  \texttt{qNoisyExpectedImprovement} \cite{balandat_botorch_2020}.

\section{Results}
\label{sec:Results}

 The selected tune parameters, motivated in Sec.\, \ref{sec:Lund_parameters}, are optimized over the ranges shown in Table \ref{table:pythia_param_table}, which are the allowable ranges in \pythia. The last column in the table gives the default \pythia tune.

\begin{table}[h!]
  \centering
  \begin{tabular}{|c|c|c|}
    \hline
    \textbf{Name in \pythia} & \textbf{Tuning range} & \textbf{\Monash}\\
    \hline
    \verb|StringZ:aLund| & $[0.0,2.0]$ & $0.68$ \\
    \hline
    \verb|StringZ:bLund| & $[0.2,2.0]$ & $0.98$\\
    \hline
    \verb|StringFlav:ProbStoUD| & $[0.0,1.0]$ & $0.217$ \\
    \hline
    \verb|StringFlav:probQQtoQ| & $[0.0,1.0]$ & $0.081$ \\
    \hline 
    \verb|TimeShower:alphaSvalue| & $[0.06,0.25]$ & $0.1365$ \\
    \hline 
    \verb|TimeShower:pTmin| & $[0.1,2.0]$ & $0.5$ \\
    \hline 
  \end{tabular}
  \caption{Selected \pythia tune parameters. The parameters are optimized over the ranges shown in the second column. The last column lists the \Monash tune in \pythia, which is the default tune.}
  \label{table:pythia_param_table}
\end{table}
Table~\ref{table:pythia_EI_all_histograms} shows the results of five BayesOpt runs using our \GPytorch implementation, where each run uses different choices of the BayesOpt hyperparameters $N_\text{BayesOpt}$, $N_\text{Sobol}$, and $N_\text{restarts}$. By definition, the optimal tune is the one with the lowest value of the objective function, $O(\mathbf{x})$, which is seen to arise from run \texttt{BO-EI5}. However, Table~\ref{table:pythia_EI_all_histograms} shows that the \Monash point gives a better fit, as indicated by its lower objective function value. We also note that the BayesOpt reaches a tune that differs from the \Monash tune. 

  Tables \ref{table:BO_qEI_results} and \ref{table:BO_qEI_results_NBO_200} show details of two different \BOTORCH runs, where each row apart from the one labeled \Monash corresponds to a different step in the Bayesian optimization. We find that the two runs arrive at essentially the same tune. Furthermore, this point is close to the tune obtained using \GPytorch, suggesting that the latter would converge to the same point had the algorithm been allowed to run for longer. Interestingly, the value of the objective function of the tune obtained with \BOTORCH is \emph{lower} than that of the \Monash point, that is, we obtain a tune that  fits the \ALEPH data better than the \Monash tune.

However, there are some caveats on the physics implications of the results.
The properties of the parton shower are set mainly by the value $\alpha_s(M_Z)$ and the shower
cut-off $p_{T,min}$.   A large high-scale value for $\alpha_s$ leads also to large values at lower
scales, suppressing the no radiation probability.   On the other hand, a large value of $p_{T,min}$
means that small values of hadron $p_T$ must come from some other mechanism.   The Lund model
provides both longitudinal and transverse momentum to hadrons and can serve as this mechanism, but only
for a certain range of parameters.   For the large values of $a_\mathrm{Lund},b_\mathrm{Lund}$ appearing in
these tunes, the average $z$ values of the string splittings are higher than those for Monash.
On the other hand, the average $z$ values are constrained by the average particle multiplicity, leading to
a tightly constrained system.
In other words, the properties of hadrons depend upon several models, but there are no clear boundaries between those models.   If the parton shower fails to produce enough transverse momentum, the hadronization model will be forced to compensate.  This could lead to a choice of parameters for the parton shower and the hadronization model that appear to be extreme but lead to a reasonable fit to the data.


Another caveat is that the predictions are computed with a statistical precision comparable to that of the \ALEPH data. One might be tempted to interpret the likelihood hyperparameter $\sigma_\text{noise}$ as a measure of the noisiness of the objective function. For our \GPytorch results, we find $\sigma_\text{noise} = 1.6$. However, this value should be interpreted with caution. A direct way to estimate the noise in the objective function is to repeatedly simulate many sets of 250,000 events at a given tune point, $\mathbf{x}$, and compute the standard deviation of $O(\mathbf{x})$.
For the \Monash point the mean of $O(\mathbf{x})$ is 3.19 and its standard deviation is 0.014. This suggests that $O(\mathbf{x})$ is a relatively smooth function of $\mathbf{x}$ despite the fact that the statistical precision of the predictions is of the same order as that of the \ALEPH data. Presumably, this is because these data are of high precision.

\begin{table*}[t]
  \centering
  \caption{\GPytorch results using all of the \ALEPH histograms and Eq.~\eqref{eq:Ox}. Each row shows the results for a BayesOpt run with the indicated hyperparameter values for $N_\text{BayesOpt}$, $N_\text{Sobol}$, and $N_\text{restart}$.}
  \begin{ruledtabular}
    \resizebox{\textwidth}{!}{%
      \begin{tabular}{rrrrrrrrrrr}
        Label           & $N_{\text{BayesOpt}}$ & $N_{\text{Sobol}}$ & $N_{\text{restarts}}$ & $O(\mathbf{\hat{x}})$ & aLund & bLund & ProbStoUD & probQQtoQ & alphaSvalue & pTmin  \\
        \hline
        \textbf{Monash} & \textbf{}       & \textbf{}          & \textbf{}              & \textbf{3.2083} & \textbf{0.68} & \textbf{0.98} & \textbf{0.217} & \textbf{0.081} & \textbf{0.1365} & \textbf{0.500} \\
        BO-EI1          & 80              & 25                 & 25                     & 7.763          & 1.388         & 1.458         & 0.019          & 0.048          & 0.133           & 0.260  \\
        BO-EI2          & 90              & 225                & 200                    & 4.082          & 0.741         & 0.646         & 0.232          & 0.077          & 0.133           & 0.344  \\
        BO-EI3          & 100             & 75                 & 25                     & 6.446          & 1.050         & 0.592         & 0.401          & 0.069          & 0.128           & 1.887  \\
        BO-EI4          & 120             & 300                & 200                    & 4.174          & 1.260         & 1.858         & 0.325          & 0.060          & 0.137           & 0.138  \\
        BO-EI5          & 180             & 300                & 5000                   & 3.789          & 1.991         & 1.873         & 0.252          & 0.084          & 0.135           & 1.207  \\
      \end{tabular}%
    }
  \end{ruledtabular}
\label{table:pythia_EI_all_histograms}
\end{table*}

\begin{table*}[htb]
  \centering
  \caption{Results of a single \texttt{BOTORCH} run using all of the \ALEPH histograms with $N_\text{BayesOpt}=100$, $N_\text{Sobol}=25$.  The first row shows the \Monash point and the optimal tune found with \texttt{BOTORCH} is in bold.}
  \begin{ruledtabular}
    \resizebox{\textwidth}{!}{%
      \begin{tabular}{lccccccccc}
        label            & $N_{\text{BayesOpt}}$ & $N_{\text{Sobol}}$ & $O(\mathbf{\hat{x}})$ & aLund & bLund & ProbStoUD & probQQtoQ & alphaSvalue & pTmin \\ 
        \hline
        \textbf{Monash}  & \textbf{}       & \textbf{}          & \textbf{3.2083} & \textbf{0.68} & \textbf{0.98} & \textbf{0.217} & \textbf{0.081} & \textbf{0.1365} & \textbf{0.500} \\
        BO-qEI1          & 100              & 25                 & 3.185          & 1.807         & 1.885         & 0.228          & 0.087          & 0.141           & 1.170  \\
        BO-qEI2          & 100              & 25                 & 3.179          & 1.777         & 1.834         & 0.198          & 0.094          & 0.141           & 1.001  \\
        BO-qEI3          & 100              & 25                 & 3.156          & 1.793         & 1.837         & 0.196          & 0.085          & 0.139           & 1.138  \\
        BO-qEI4          & 100              & 25                 & 3.153          & 1.904         & 1.867         & 0.209          & 0.094          & 0.138           & 0.999  \\
        BO-qEI5          & 100              & 25                 & 3.152          & 2.000         & 1.841         & 0.204          & 0.085          & 0.143           & 1.326  \\
        BO-qEI6          & 100              & 25                 & 3.145          & 2.000         & 1.813         & 0.200          & 0.072          & 0.136           & 1.118  \\
        BO-qEI7          & 100              & 25                 & 3.139          & 1.825         & 1.935         & 0.203          & 0.087          & 0.139           & 0.987  \\
        BO-qEI8          & 100              & 25                 & 3.131          & 2.000         & 1.938         & 0.209          & 0.083          & 0.135           & 0.748  \\
        BO-qEI9          & 100              & 25                 & 3.130          & 1.842         & 1.853         & 0.221          & 0.090          & 0.138           & 0.845  \\
        BO-qEI10         & 100              & 25                 & 3.115          & 1.815         & 1.840         & 0.198          & 0.081          & 0.136           & 0.916  \\
        BO-qEI11         & 100              & 25                 & 3.102          & 2.000         & 1.870         & 0.220          & 0.079          & 0.136           & 0.933  \\
        BO-qEI12         & 100              & 25                 & 3.095          & 1.893         & 1.867         & 0.221          & 0.078          & 0.138           & 0.917  \\
        BO-qEI13         & 100              & 25                 & 3.073          & 1.984         & 1.935         & 0.217          & 0.087          & 0.138           & 0.938  \\
        BO-qEI14         & 100              & 25                 & 3.054          & 1.937         & 1.875         & 0.196          & 0.080          & 0.138           & 1.060  \\
        BO-qEI15         & 100              & 25                 & 3.050          & 2.000         & 1.640         & 0.207          & 0.082          & 0.142           & 1.649  \\
        BO-qEI16         & 100              & 25                 & 3.049          & 2.000         & 1.594         & 0.215          & 0.078          & 0.137           & 1.398  \\
        BO-qEI17         & 100              & 25                 & 3.042          & 1.945         & 1.888         & 0.211          & 0.076          & 0.139           & 1.183  \\
        BO-qEI18         & 100              & 25                 & 3.034          & 1.875         & 1.588         & 0.176          & 0.085          & 0.139           & 1.465  \\
        BO-qEI19         & 100              & 25                 & 3.033          & 2.000         & 1.837         & 0.232          & 0.075          & 0.139           & 1.235  \\
        BO-qEI20         & 100              & 25                 & 3.029          & 1.862         & 1.895         & 0.210          & 0.077          & 0.137           & 0.986  \\
        BO-qEI21         & 100              & 25                 & 3.023          & 1.939         & 1.881         & 0.225          & 0.085          & 0.140           & 1.139  \\
        BO-qEI22         & 100              & 25                 & 3.007          & 1.818         & 1.807         & 0.204          & 0.076          & 0.138           & 1.075  \\
        BO-qEI23         & 100              & 25                 & 3.006          & 1.901         & 1.768         & 0.197          & 0.078          & 0.139           & 1.333  \\
        BO-qEI24         & 100              & 25                 & 3.006          & 1.886         & 1.842         & 0.223          & 0.080          & 0.138           & 1.128  \\
        BO-qEI25         & 100              & 25                 & 3.000          & 1.874         & 1.776         & 0.214          & 0.074          & 0.138           & 1.166  \\
        BO-qEI26         & 100              & 25                 & 2.986          & 1.918         & 1.697         & 0.214          & 0.076          & 0.137           & 1.292  \\
        BO-qEI27         & 100              & 25                 & 2.946          & 1.987         & 1.650         & 0.204          & 0.079          & 0.140           & 1.508  \\
        BO-qEI28         & 100              & 25                 & 2.939          & 1.907         & 1.643         & 0.223          & 0.078          & 0.139           & 1.444  \\
        \textbf{BO-qEI29} & \textbf{100}     & \textbf{25}       & \textbf{2.875} & \textbf{1.839} & \textbf{1.587} & \textbf{0.205} & \textbf{0.077} & \textbf{0.139} & \textbf{1.533} \\
      \end{tabular}%
    }
  \end{ruledtabular}
\label{table:BO_qEI_results}
\end{table*}

\begin{figure*}[ht]
    \centering
    \includegraphics[width=\textwidth]{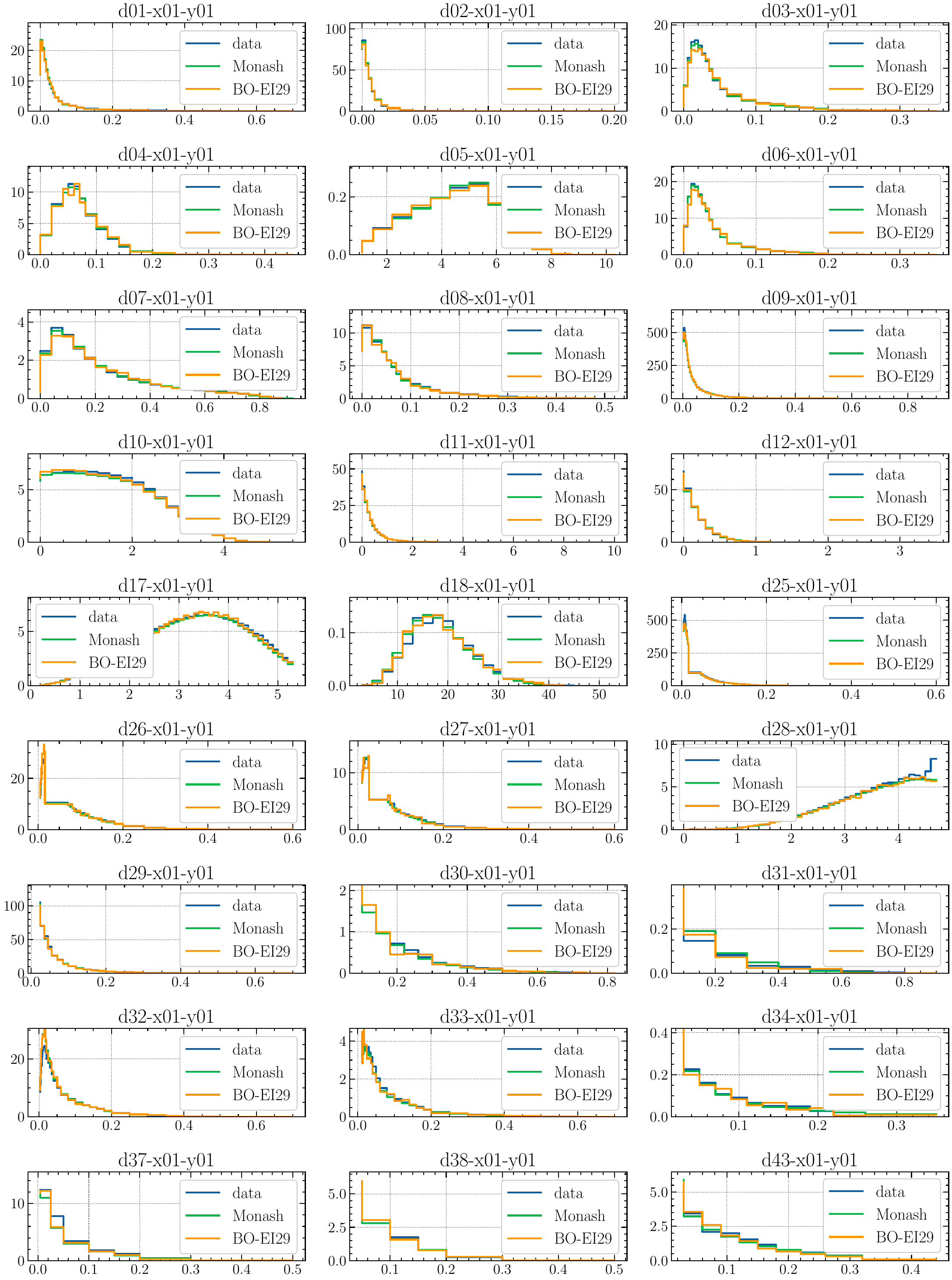}
    \caption{The BO EI tune results using \GPytorch for label ``BO-EI4" in Table \ref{table:pythia_EI_all_histograms}}
    \label{fig:EI4_pythia}
\end{figure*}

\begin{figure*}[ht]
    \centering
    \includegraphics[width=\textwidth]{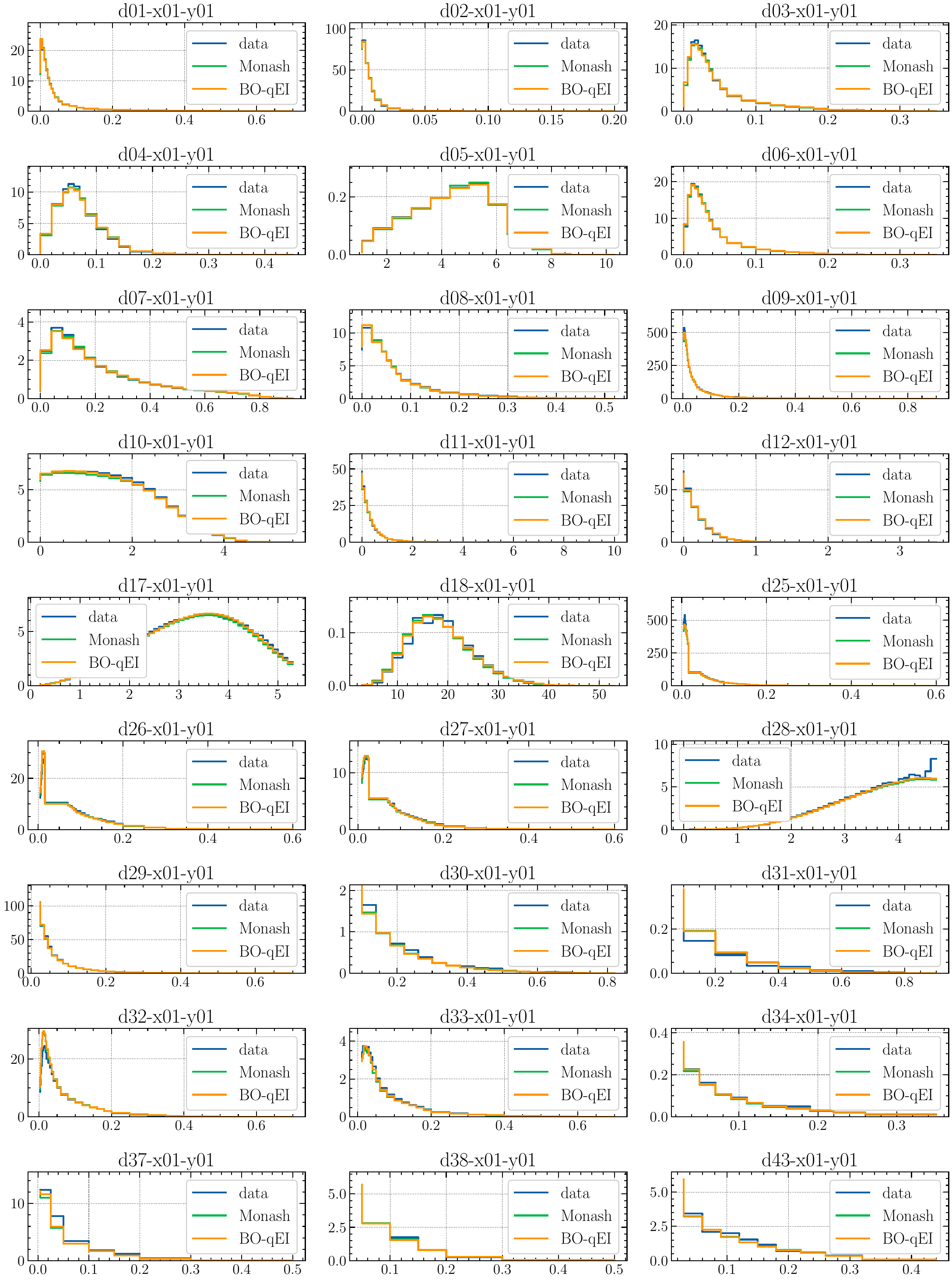}
    \caption{\texttt{BOTORCH} results on all of the histograms and Eq. \ref{eq:Ox} with $N_{BO}=100, N_{Sobol}=25$, corresponding to the bolded row in Table \ref{table:BO_qEI_results}}
\label{fig:BOTorch_qEI29}
\end{figure*}

\clearpage
\begin{table*}[htb]
  \centering
  \footnotesize
  \caption{Results of a single \texttt{BOTORCH} run using all of the \ALEPH histograms with $N_\text{BayesOpt}=200$, $N_\text{Sobol}=30$.  The first row shows the \Monash point and the optimal tune found with \texttt{BOTORCH} is in bold.}
  \begin{ruledtabular}
    \resizebox{\textwidth}{!}{%
      \begin{tabular}{rrrrrrrrrrr}
        \toprule
        label        & $N_{\rm BaysOpt}$ & $N_{\rm Sobol}$ & $O(\mathbf{\hat{x}})$ & aLund & bLund & ProbStoUD & probQQtoQ & alphaSvalue & pTmin \\
        \midrule
        \textbf{Monash}    & \textbf{}  & \textbf{}  & \textbf{3.2083} & \textbf{0.68} & \textbf{0.98} & \textbf{0.217} & \textbf{0.081} & \textbf{0.1365} & \textbf{0.500} \\
        BO-qEI-0           & 200 & 30 & 2.944 & 1.639 & 1.573 & 0.210 & 0.083 & 0.139 & 1.184 \\
        BO-qEI-1           & 200 & 30 & 2.941 & 1.757 & 1.557 & 0.204 & 0.086 & 0.139 & 1.437 \\
        BO-qEI-2           & 200 & 30 & 2.931 & 1.929 & 1.559 & 0.215 & 0.077 & 0.140 & 1.771 \\
        BO-qEI-3           & 200 & 30 & 2.929 & 1.820 & 1.619 & 0.207 & 0.081 & 0.138 & 1.208 \\
        BO-qEI-4           & 200 & 30 & 2.928 & 1.885 & 1.556 & 0.198 & 0.080 & 0.139 & 1.637 \\
        BO-qEI-5           & 200 & 30 & 2.925 & 1.807 & 1.584 & 0.204 & 0.082 & 0.139 & 1.418 \\
        BO-qEI-6           & 200 & 30 & 2.922 & 1.796 & 1.573 & 0.207 & 0.077 & 0.139 & 1.489 \\
        BO-qEI-7           & 200 & 30 & 2.920 & 1.842 & 1.524 & 0.216 & 0.081 & 0.139 & 1.655 \\
        BO-qEI-8           & 200 & 30 & 2.919 & 1.802 & 1.549 & 0.218 & 0.080 & 0.139 & 1.512 \\
        BO-qEI-9           & 200 & 30 & 2.915 & 1.783 & 1.543 & 0.214 & 0.082 & 0.139 & 1.568 \\
        BO-qEI-10          & 200 & 30 & 2.914 & 1.809 & 1.570 & 0.213 & 0.085 & 0.140 & 1.446 \\
        BO-qEI-11          & 200 & 30 & 2.913 & 1.807 & 1.567 & 0.198 & 0.085 & 0.139 & 1.385 \\
        BO-qEI-12          & 200 & 30 & 2.913 & 1.743 & 1.553 & 0.210 & 0.085 & 0.138 & 1.283 \\
        BO-qEI-13          & 200 & 30 & 2.912 & 1.842 & 1.555 & 0.202 & 0.083 & 0.140 & 1.549 \\
        BO-qEI-14          & 200 & 30 & 2.911 & 1.779 & 1.568 & 0.213 & 0.082 & 0.138 & 1.380 \\
        BO-qEI-15          & 200 & 30 & 2.911 & 1.855 & 1.559 & 0.201 & 0.085 & 0.139 & 1.540 \\
        BO-qEI-16          & 200 & 30 & 2.910 & 1.841 & 1.582 & 0.212 & 0.087 & 0.139 & 1.479 \\
        BO-qEI-17          & 200 & 30 & 2.909 & 1.838 & 1.584 & 0.217 & 0.083 & 0.138 & 1.408 \\
        BO-qEI-18          & 200 & 30 & 2.902 & 1.814 & 1.582 & 0.201 & 0.085 & 0.139 & 1.424 \\
        BO-qEI-19          & 200 & 30 & 2.901 & 1.788 & 1.571 & 0.204 & 0.080 & 0.139 & 1.446 \\
        BO-qEI-20          & 200 & 30 & 2.898 & 1.887 & 1.550 & 0.211 & 0.083 & 0.140 & 1.636 \\
        BO-qEI-21          & 200 & 30 & 2.896 & 1.756 & 1.564 & 0.216 & 0.080 & 0.140 & 1.415 \\
        BO-qEI-22          & 200 & 30 & 2.895 & 1.846 & 1.594 & 0.206 & 0.080 & 0.139 & 1.444 \\
        BO-qEI-23          & 200 & 30 & 2.893 & 1.840 & 1.570 & 0.206 & 0.087 & 0.139 & 1.433 \\
        BO-qEI-24          & 200 & 30 & 2.892 & 1.870 & 1.546 & 0.212 & 0.077 & 0.139 & 1.613 \\
        BO-qEI-25          & 200 & 30 & 2.892 & 1.824 & 1.555 & 0.219 & 0.081 & 0.139 & 1.538 \\
        BO-qEI-26          & 200 & 30 & 2.891 & 1.787 & 1.580 & 0.208 & 0.081 & 0.138 & 1.318 \\
        BO-qEI-27          & 200 & 30 & 2.888 & 1.793 & 1.570 & 0.214 & 0.081 & 0.139 & 1.321 \\
        BO-qEI-28          & 200 & 30 & 2.886 & 1.830 & 1.612 & 0.215 & 0.082 & 0.139 & 1.455 \\
        \textbf{BO-qEI-29} & \textbf{200} & \textbf{30} & \textbf{2.845} & \textbf{1.836} & \textbf{1.564} & \textbf{0.208} & \textbf{0.079} & \textbf{0.139} & \textbf{1.549} \\
        \bottomrule
      \end{tabular}%
    }
  \end{ruledtabular}
\label{table:BO_qEI_results_NBO_200}
\end{table*}

\section{Conclusions}
\label{sec:conclusions}
Bayesian optimization has been used to tune six parameters that are most relevant for modeling hadronization in \pythia. Two different implementations of BayesOpt (\GPytorch and \BOTorch) yield essentially the same tune point, which differs from the \Monash tune. Moreover, we find that the tune obtained with \BOTORCH fits the \ALEPH data better than does the \Monash tune. However, in drawing conclusions about the new tune, it is important to take note of the caveats described above. Through extensive investigation, we find that the most important factor in producing a good universal \pythia tune is to use all the histograms in the \ALEPH data.

All results are reproducible using the resources at \url{https://github.com/AliAlkadhim/BayesOptPythia/}.

\begin{acknowledgments}
This document was prepared using the resources of the Fermi National Accelerator Laboratory (Fermilab), a U.S. Department of Energy, Office of Science, Office of High Energy Physics HEP User Facility. Fermilab is managed by FermiForward Discovery Group, LLC, acting under Contract No. 89243024CSC000002. This work is supported in part by the US Department of Energy under Award No. DE-SC0010102 and in part by the University Research Association (URA)
Visiting Scholars Program Award No. 23-S-01. We are thankful to Pushpalatha Bhat and Tony Menzo for their insightful feedback on this project. Special thanks to the Fermilab LHC Physics Center (LPC) for providing access to resources and facilities that were instrumental to the completion of this work.

\end{acknowledgments}


\appendix
\section{Reproducibility}
\label{sec:reproducibility}
All of the code, data, and execution environment needed to reproduce our toy‑problem studies and the \pythia tune are publicly available.

\subsection{Repository and Code}
The complete analysis is hosted at:
\begin{center}
  \url{https://github.com/AliAlkadhim/BayesOptPythia}
\end{center}
Key scripts and notebooks live under \texttt{BayesOpt/src/}:
\begin{itemize}
  \item \texttt{main\_toy.py} — single toy‑problem Bayesian‐optimization run.  
  \item \texttt{main\_toy\_all\_hyperparams.py} — full 108‑experiment hyperparameter scan.  
  \item \texttt{validate\_pythia\_EI.py} — \textsc{Pythia\,8} Bayesian‐optimization tune.  
  \item \texttt{post\_processing\_toy.ipynb}, \texttt{post\_processing\_pythia.ipynb} — Jupyter notebooks to regenerate all figures and tables.  
\end{itemize}

\subsection{Docker Environment}
To avoid dependency issues, we provide a Docker image with all required software 
such as \pythia version 8.309, \rivet version 3.1.9, \yoda version 1.9.9, \HEPMC version 3.02.06, \LHAPDF version 6.5.4, \GPytorch, \BOTORCH, etc.  After installing Docker, pull and run:
\begin{verbatim}
docker pull alialkadhim/pythia_sbi_tune:latest
docker run -v $PWD:$PWD \
           -w $PWD \
           -p 8889:8889 \
           -it alialkadhim/pythia_sbi_tune:latest
\end{verbatim}
Inside the container:
\begin{verbatim}
source setup.sh
jupyter-lab --ip 0.0.0.0 --port 8889 --allow-root &
\end{verbatim}
Then open the printed URL in your browser.

\subsection{Running the Experiments}
\paragraph{Toy Problems}
\begin{enumerate}
  \item Edit hyperparameters in \texttt{BayesOpt/src/configs.py}.  
  \item Run a single toy run:
    \begin{verbatim}
    python BayesOpt/src/main_toy.py
    \end{verbatim}
  \item To reproduce the full hyperparameter scan:
    \begin{verbatim}
    python BayesOpt/src/main_toy_all_hyperparams.py
    \end{verbatim}
  \item Generate diagnostic plots by executing \texttt{post\_processing\_toy.ipynb}.  
\end{enumerate}

\paragraph{\textsc{Pythia\,8} Tune}
\begin{enumerate}
  \item Configure the tune parameters in \texttt{configs.py}.  
  \item Launch the tune:
    \begin{verbatim}
    python BayesOpt/src/validate_pythia_EI.py
    \end{verbatim}
  \item Produce comparison plots and uncertainty envelopes via \texttt{post\_processing\_pythia.ipynb}.  
\end{enumerate}

\subsection{Computational Resources}
A full \pythia tune of 6 parameters \(\times\) \(250,000\) events per point) takes around 374 seconds on a machine with the following specifications: 8th Generation Intel Core i7-855OU Processor, 16 GB DDR4-2400 SDRAM (2 x 8GB). For a reasonable run of $N_\text{BaysOpt}=70$ optimization this amounts to 7.3 hours. The majority of this time is taken up by \rivet. Although we consider a very large number of restarts in our EI \GPytorch experiments, the time taken by the optimization is small in comparison to the time required for \pythia and \rivet.



\section{Description of \LEP Data}
\begin{table}[htb]
\centering
\begin{tabular}{l|l}
\hline
\textbf{histogram label} & \textbf{title} \\
\hline
d01-x01-y01 & Sphericity, $S$ (charged) \\
d02-x01-y01 & Aplanarity, $A$ (charged) \\
d03-x01-y01 & 1-Thrust, $1-T$ (charged) \\
d04-x01-y01 & Thrust minor, $m$ (charged) \\
d05-x01-y01 & Two-jet resolution variable, $Y_3$ (charged) \\
d06-x01-y01 & Heavy jet mass (charged) \\
d07-x01-y01 & $C$ parameter (charged) \\
d08-x01-y01 & Oblateness, $M - m$ (charged) \\
d09-x01-y01 & Scaled momentum, $x_p = |p|/|p_\text{beam}|$ (charged) \\
d10-x01-y01 & Rapidity w.r.t. thrust axes, $y_T$ (charged) \\
d11-x01-y01 & In-plane $p_T$ in GeV w.r.t. sphericity axes (charged) \\
d12-x01-y01 & Out-of-plane $p_T$ in GeV w.r.t. sphericity axes (charged) \\
d17-x01-y01 & Log of scaled momentum, $\log(1/x_p)$ (charged) \\
d18-x01-y01 & Charged multiplicity distribution \\
d19-x01-y01 & Mean charged multiplicity \\
d20-x01-y01 & Mean charged multiplicity for rapidity $|Y| < 0.5$ \\
d21-x01-y01 & Mean charged multiplicity for rapidity $|Y| < 1.0$ \\
d22-x01-y01 & Mean charged multiplicity for rapidity $|Y| < 1.5$ \\
d23-x01-y01 & Mean charged multiplicity for rapidity $|Y| < 2.0$ \\
d25-x01-y01 & $\pi^\pm$ spectrum \\
d26-x01-y01 & $K^\pm$ spectrum \\
d27-x01-y01 & $p$ spectrum \\
d28-x01-y01 & $\gamma$ spectrum \\
d29-x01-y01 & $\pi^0$ spectrum \\
d30-x01-y01 & $\eta$ spectrum \\
d31-x01-y01 & $\eta^\prime$ spectrum \\
d32-x01-y01 & $K^0$ spectrum \\
d33-x01-y01 & $\Lambda^0$ spectrum \\
d34-x01-y01 & $\Xi^-$ spectrum \\
d35-x01-y01 & $\Sigma^\pm(1385)$ spectrum \\
d36-x01-y01 & $\Xi^0(1530)$ spectrum \\
d37-x01-y01 & $\rho$ spectrum \\
d38-x01-y01 & $\omega(782)$ spectrum \\
d39-x01-y01 & $K^{*0}(892)$ spectrum \\
d40-x01-y01 & $\phi$ spectrum \\
d43-x01-y01 & $K^{*\pm}(892)$ spectrum \\
d44-x01-y02 & Mean $\pi^0$ multiplicity \\
d44-x01-y03 & Mean $\eta$ multiplicity \\
d44-x01-y04 & Mean $\eta^\prime$ multiplicity \\
d44-x01-y05 & Mean $K_S + K_L$ multiplicity \\
d44-x01-y06 & Mean $\rho^0$ multiplicity \\
d44-x01-y07 & Mean $\omega(782)$ multiplicity \\
d44-x01-y08 & Mean $\phi$ multiplicity \\
d44-x01-y09 & Mean $K^{*\pm}$ multiplicity \\
d44-x01-y10 & Mean $K^{*0}$ multiplicity \\
d44-x01-y11 & Mean $\Lambda$ multiplicity \\
d44-x01-y12 & Mean $\Sigma$ multiplicity \\
d44-x01-y13 & Mean $\Xi$ multiplicity \\
d44-x01-y14 & Mean $\Sigma(1385)$ multiplicity \\
d44-x01-y15 & Mean $\Xi(1530)$ multiplicity \\
d44-x01-y16 & Mean $\Omega^\mp$ multiplicity \\
\hline
\end{tabular}
\caption{Histogram labels and their corresponding titles for the \ALEPH data.}
\label{table:all_hists}
\end{table}

\clearpage
\bibliographystyle{unsrt}
\bibliography{references}
\end{document}